\documentclass{iopart}
\usepackage[pdftex]{graphicx}
\usepackage[latin1]{inputenc}
  \expandafter\let\csname equation*\endcsname\relax
  \expandafter\let\csname endequation*\endcsname\relax 
\usepackage{amsmath}
\usepackage{iopams}

\newcommand{\whz}{\sqrt{\rm Hz}}

\begin{document}

\title[Phasemeter core for intersatellite laser heterodyne interferometry]{Phasemeter core for intersatellite laser heterodyne interferometry: modelling, simulations and experiments}

\author{Oliver~Gerberding, Benjamin~Sheard, Iouri~Bykov, Joachim~Kullmann, Juan~Jose~Esteban~Delgado, Karsten~Danzmann and Gerhard~Heinzel}
\address{Max Planck Institute for Gravitational Physics, and Institute for Gravitational Physics of the Leibniz Universität Hannover, Callinstrasse 38, 30167 Hannover, Germany}

\ead{oliver.gerberding@aei.mpg.de}

\begin{abstract} 
Inter satellite laser interferometry is a central component of future
space-borne gravity instruments like LISA, eLISA, NGO and future geodesy missions.
The inherently small laser wavelength allows to measure distance variations with extremely high precision
by interfering a reference beam with a measurement beam.
The readout of such interferometers is often based on tracking phasemeters, able
to measure the phase of an incoming beatnote with high precision over a wide
range of frequencies. The implementation of such phasemeters is based on all digital
phase-locked loops, hosted in FPGAs. Here we present a precise model of an all
digital phase locked loop that allows to design such a readout algorithm and we
support our analysis by numerical performance measurements and experiments with analog signals.
\end{abstract}



%
%
%
%

\section{Introduction}
The Laser Interferometer Space Antenna (LISA) is a space borne observatory for gravitational waves
in the frequency range of $0.1\, \textrm{mHz}$ to $1\, \textrm{Hz}$ \cite{Danzmann2003}. 
LISA will detect gravitational waves by measuring the variation of the light travel time between free-floating
test-masses with millions kilometre separation. 
Heterodyne laser interferometry is used to convert the path length variations into phase shifts of the heterodyne
beat note (2 ... 25 MHz), which is then detected by a photodiode and an electronic phasemeter.
The measurement needs to be performed with a noise level of
the order of $\mu \textrm{cycle} / \sqrt{\textrm{Hz}}$.

\begin{figure}
   \centering
   	\includegraphics[width=0.85\textwidth]{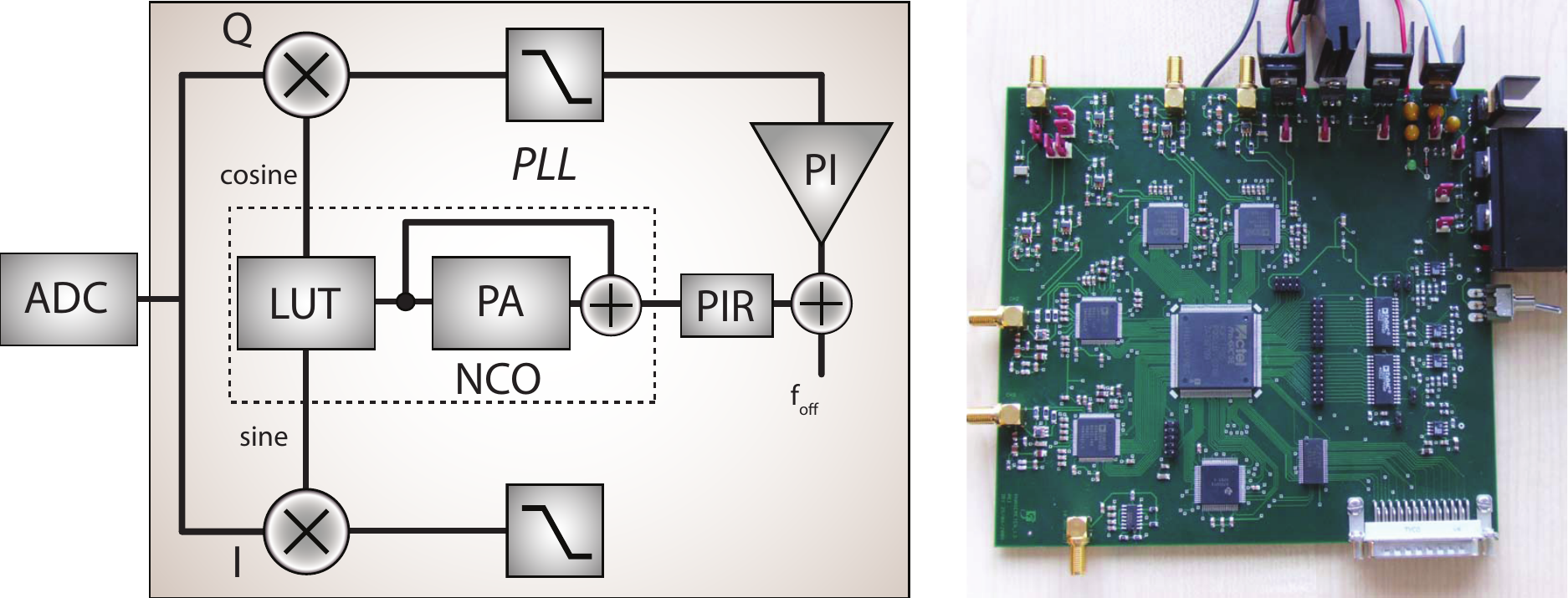}
   		\caption[figure1]{\label{figure1} The left side shows the basic phasemeter topology, where the incoming signal is digitised and then mixed in phase and in quadrature with a digital reference signal. The Quadrature information is filtered and fed into a controller (PI), determining a frequency actuation signal in the so called phase increment register (PIR). The phase increment is then fed into a numerically controlled oscillator (NCO), where it is integrated in the phase accumulator (PA) and then converted to a sine and cosine in the look-up table (LUT). The right side shows a prototype LISA phasemeter \cite{Bykov2009}. The FPGA of such a system was used to perform the digital simulations.}
\end{figure}

The readout system, or phasemeter, for these interferometers is implemented by digitising the heterodyne signals and determining the phase using an IQ demodulation system \cite{Shaddock2006,Bykov2009} implemented in an FPGA. Due to the high initial phase noise measured by each interferometer and also the continuously varying Doppler shifts in the LISA constellation, this digital IQ demodulation is embedded in a closed-loop phase and frequency tracking system, a phase-locked loop (PLL) or more specifically an all digital phase-locked loop (ADPLL). The phasemeter needs to be able to track signals between 2 and 25 MHz with a precision of $2\pi \mu \textrm{rad}/\sqrt{\textrm{Hz}}$.
An overview of the phasemeter structure and a prototype implementation is shown in figure \ref{figure1}.

In this article we present a detailed analysis of the digital core of the phasemeter and the phase tracking algorithm, based on which one can design and optimize the readout for inter satellite interferometers like LISA or future geodesy missions like GRACE Follow-On \cite{Sheard2012}. 
Even tough the original LISA design is currently not considered any more, various variants of the concept are studied, including the currently proposed evolved LISA (eLISA) \cite{Danzmann2013}. 
Since the original LISA design has comparable requirements for the phasemeter the analysis presented in this paper is aimed at this concept. Therefore the analysis can be easily adapted also for other LISA-like missions, to all of which we will refer to only as LISA in the following.

For our analysis we use a combination of analytic modelling and numeric measurements, which are performed by directly using very high speed integrated circuit hardware description language (VHDL) code running on phasemeter prototypes. This also allows us to test the algorithm properties under realistic conditions, by either using the hardware to generate realistic signals or by directly measuring analog signals. 
The article includes a linearised model of the PLL, a model for noise introduced by quantization effects, an estimate of the phasemeter linearity and the results of the digital signal measurements. In addition we present a test of the linearity performance of our prototypes using analog signals.

\section{ADPLL Model}
\subsection{Scaling}
The PLL is implemented using integer arithmetic, where each value is represented by X bits and one can use different scalings to map these to numbers. In this article we choose to scale each integer by $2^{-X}$, which leads to the following ranges for signed and unsigned numbers:
\begin{equation}
	\begin{split}
		 \mathbf{ -0.5 \leq} & \left[ \frac{-2^{X-1}}{2^X}  \leq \textrm{\bf ~signed~} \leq  \frac{2^{X-1}-1}{2^X}\right]  \mathbf{ < 0.5}  \\
		 \mathbf{ 0 \leq} &  \left[ ~~~ \frac{0}{2^X} ~~  \leq \textrm{\bf unsigned} \leq ~ \frac{2^{X}-1}{2^X} ~ \right]  \mathbf{ < 1} 
	\end{split}
	\label{eq:sclaing}
\end{equation}
Numbers with this scaling have no units, since they only represents values in the digital computation, only appropriate further scaling maps them to real physical quantities with units, as done in the following.

%
\subsection{Linearised model}
The specific linear model presented here is a modification of well known ADPLL models described by Gardner \cite{Gardner2005}.
Figure 2 shows the block diagram of the model, which considers phase as the quantity that is sensed and actuated. 
The signals and blocks shown are described in the following.

\begin{figure}
  \centering
  \includegraphics[width=0.85\textwidth]{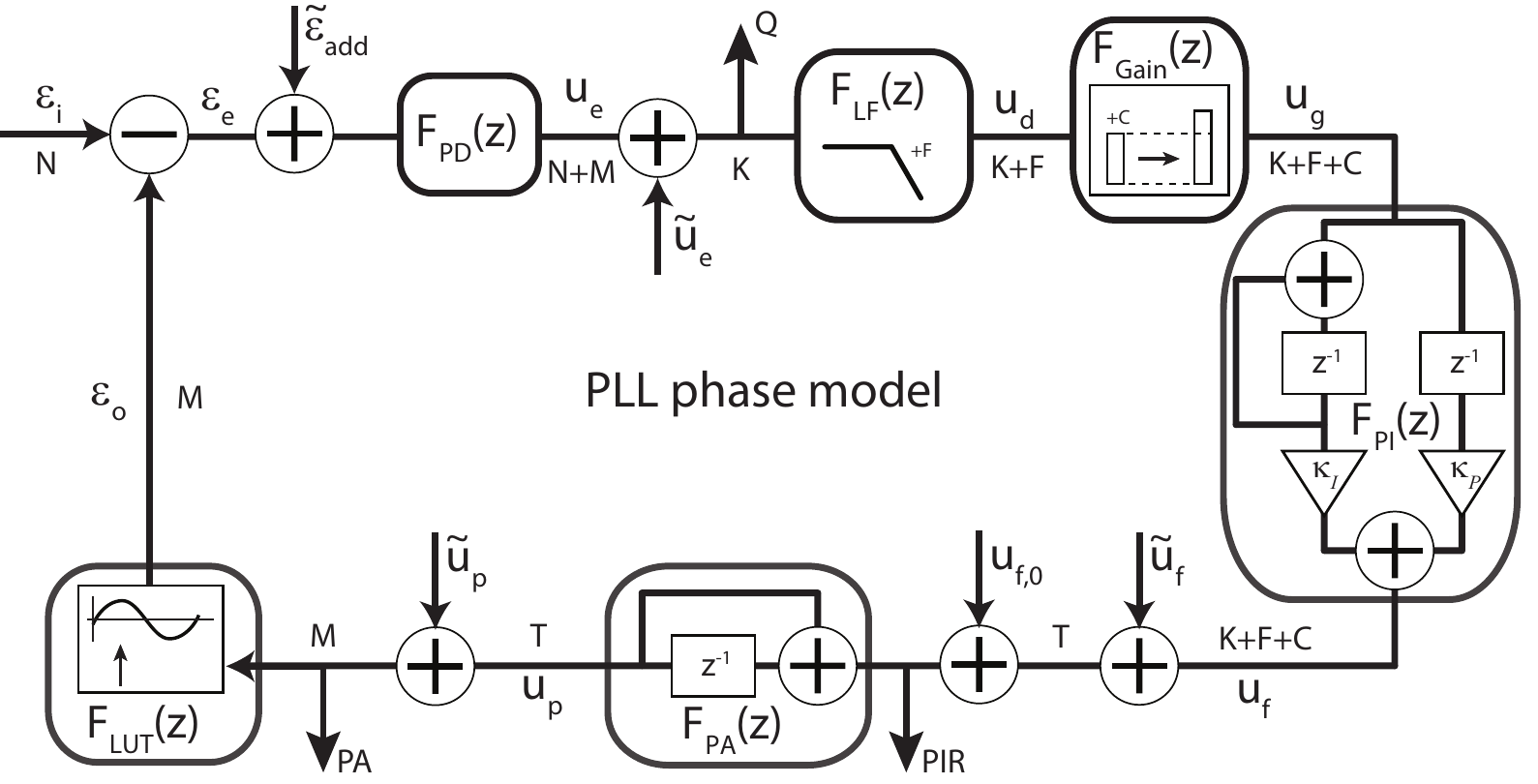}
   		\caption[figure2]{\label{figure2} Block diagram of the linearised PLL model. Included are
   		bitlength indicators (N,M,K,F,C,T), possible truncation noise additions ($\widetilde{u}_e$, $\widetilde{u}_f$, $\widetilde{u}_p$),
   		input additive noise $\widetilde{\varepsilon}_{\textrm{add}}$ and markers for signal readout points(Q, PA, PIR).
   		Not shown here are the amplitude readout and the additional computation delay transfer function. The controller and the low pass
   		filter can be implemented according to signal and requirement specifications. The input and output of the LUT is kept at
   		an equal bitlength in the linear design of the model. (A larger output does not lead to phase noise
   		improvement, since the phase information is already lost before the LUT)}
\end{figure}

Assuming an input signal with a peak amplitude $V_{\text{in}}$ in Volts and a maximum peak-to-peak voltage range of the analog-to-digital converter of $V_{\text{p-p}}$, the digitized input signal $i[n]$ \footnote{In the following we specify for a quantity $x$ also its unit $[x]$ and range $(\textrm{min} < x < \textrm{max})$ } is
\begin{equation}
\begin{split}
		i[n] 
		&= \frac{  V_{\text{in}}}  {V_{\text{p-p}}} \cdot \sin{\left( \omega_0 n  + \varepsilon_{i}[n] \right)}
		= A \cdot \sin{\left( \omega_0 n  + \varepsilon_{i}[n] \right)}, \\
				 \left[i[n] \right] &= 1; ~ \left(-0.5 \leq i[n] <0.5 \right) \\
				 \left[A \right] &= 1; ~ \left(0 \leq A <0.5 \right)
		\end{split}
	\label{incoming1}
\end{equation}
with $\varepsilon_i[n]$ as phase, the signal of interest, and $\omega_0$ ($[\omega_0]= \textrm{cycle}$) as value corresponding to the beatnote frequency $f_0$ ($[f_0]= \textrm{Hz}$) in a digital system sampled by a sampling frequency $f_s$. 
This leads to time steps between two samples $n$ and $n+1$ of $\tau_s = 1/f_s$.
 One should note here that $\omega_0$ is only necessary for the initial loop acquisition and not for the linear model, it is included here to keep the resemblance to the actual signals inside the PLL.
Additional terms for additive noise and additional tones are not included here.
The output of the numerically controlled oscillator $o[n]$ is described as
\begin{equation}
	\begin{split}
		o[n] &= \frac{1}{2} \cdot \cos{\left( \omega_0 n  + \varepsilon_{o}[n] \right)}, \\
						 \left[o[n] \right] &= 1; ~ \left(-0.5 \leq o[n] <0.5 \right) \\
	\end{split}
	\label{incoming2}
\end{equation}
with $\varepsilon_{o}[n]$ as NCO output phase, the current PLL reference. The ideal error signal of the loop
 ($\varepsilon_e = \varepsilon_{i}[n] - \varepsilon_{o}[n]$) is not directly accessible by arithmetic operations,
 therefore it is approximated by multiplying both signals to compute an error signal $u_e [n]$,
\begin{equation}
\begin{split}
 u_e [n] =& i[n]  \cdot o[n] \\
	u_e [n] =& 	 \frac{A}{4} \cdot \big[  \sin{\left(  \varepsilon_{e}[n] \right)}  + 
		 \sin{\left(  2 \omega_0 n  + \varepsilon_{i}[n] + \varepsilon_{o}[n] \right)}     \big].
\end{split}
	\label{eq:mixer1}
\end{equation}
At this point two linearisations are introduced to complete the linear model. First we assume a 
small phase error $\left( \varepsilon_{e}[n] \ll 1 \right)$, which implies the loop is locked with sufficient loop gain,
 and second, we assume a suppression of the 
second harmonic term by appropriate filtering (this also includes the suppression of additional tones).
This simplifies equation (\ref{eq:mixer1}) to
\begin{equation}
\begin{split}
	u_e [n] & \approx 	 \frac{A}{4} \left(  \varepsilon_{i}[n] - \varepsilon_{o}[n] \right) = 
		 \frac{A}{4} \left(  \varepsilon_{e}[n] \right), \\
		 \left[u_e[n] \right] &= \text{rad} .
	\end{split}
	\label{mixer2}
\end{equation}
The phase detector can now be described with a linear transfer function including the
signal amplitude as part of its gain,
\begin{equation}
\begin{split}
	F_{\textrm{PD}} (z) &= \frac{u_e(z)}{\varepsilon_{i}(z) -
	\varepsilon_{o}(z)} =    \frac{u_e(z)}{\varepsilon_{e}(z)}	
	= \frac{A}{4}, \\
		 \left[F_{\textrm{PD}} (z) \right] &= 1 .
	\end{split}
	\label{mixer3}
\end{equation}
A generic low pass filter follows the phasedetector and provides the suppression of higher harmonics. 
The design and implementation of this filter depends on the exact loop design and should be adapted accordingly. A more detailed discussion is shown in the non-linearity section.
\begin{equation}
\begin{split}
	F_{\text{LF}} (z) &= \frac{u_d(z)}{u_e(z)}, \\
		 \left[F_{\text{LF}} (z) \right] &= 1 .
	\end{split}
	\label{loopfilter}
\end{equation}
The open-loop gain of the PLL is determined by a controller, for example a simple proportional-integral controller.
For our implementation the full loop model shows that an overall gain reduction in the loop is necessary to achieve a stable
condition. Therefore we include a constant gain reduction before the servo, to allow the system to operate at the correct
range and to prevent any overflows in the digital accumulators, where the fixed-point arithmetic is performed.
For convenience we use bit shifting, adding a number of C bits from the left to the signal leads to
\begin{equation}
\begin{split}
	F_{\text{Gain}} (z) &= \frac{u_g(z)}{u_d(z)} = 2^{-C},\\
		 \left[F_{\text{Gain}} (z) \right] &= 1 .
	\end{split}
	\label{TFGain}
\end{equation}
In the servo amplifier the desired bandwidth and loop response can then be set, by tuning
the $\kappa_p$ and $\kappa_i$ values:
\begin{equation}
\begin{split}
	F_{\text{PI}} (z) &= \frac{u_f(z)}{u_g(z)} = \kappa_p + \kappa_i\frac{z^{-1}}{1-z^{-1}},\\
		 \left[F_{\text{PI}} (z) \right] &= \frac{\text{cycle}}{\text{s} *\text{rad}} .
	\end{split}
	\label{TFPI}
\end{equation}
The frequency signal $u_f$ is now representing the frequency of the NCO and, assuming the PLL is 
locked, also allows to determine the phase of the incoming signal. 
The register containing this value
is also denoted as phase increment register (PIR). For lock acquisition, this value must be pre-set close
to the incoming frequency.

By accumulating the PIR value in a register called phase accumulator (PA) the phase driving the NCO is generated:
\begin{equation}
\begin{split}
	F_{\text{PA}} (z) &= \frac{u_p(z)}{u_f(z)} = \frac{z^{-1}}{1-z^{-1}}\\
		 \left[F_{\text{PA}} (z) \right] &=\text{s} .
	\end{split}
	\label{TFPA}
\end{equation}
This phase is then fed into a sine and cosine look-up table to generate the local oscillator. 
In the loop, this operation is
described by the transfer function
\begin{equation}  
\begin{split}
		F_{\text{LUT}} &= \frac{\varepsilon_o(z)}{u_p(z)} = 2\pi \\
		[F_{\text{LUT}}] &= \frac{\text{rad}}{\text{cycle}}.
			\end{split}
	\label{luttf1}
\end{equation}
One additional element not yet included are the delays of the signal processing. These delays become important
for high bandwidth and they can directly be computed from the number of registers used in the loop logic. For
a total delay of D clock cycles they are included as $z^{-\text{D}}$.
If parts of the loop are running at slower frequencies, the delays should be scaled according to the
sampling rate of the signal.

\begin{figure}
   \centering
   	\includegraphics[width=1\textwidth]{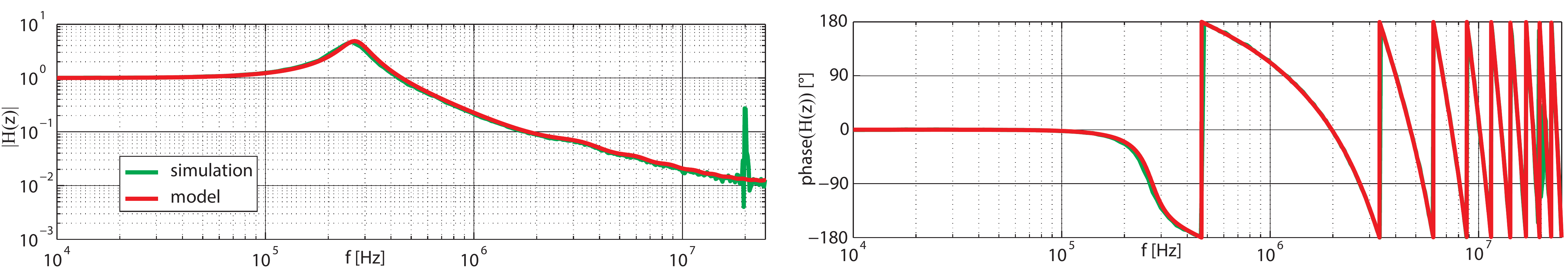}
   		\caption[figure3]{\label{figure3}
	Example of a closed-loop PLL transfer function $H(z)$ from a simulation and the linear model. Both curves are in very good agreement, save for the second harmonic, present at 20MHz.}
\end{figure}

Continuing, the above results in an open loop transfer function $G(z)$, which allows to
to determine loop stability, noise suppression and suppression of higher harmonics:

\begin{equation}
G(z) = \frac{\varepsilon_o(z)}{\varepsilon_e(z)} =
\frac{A \pi}{2} 
\cdot
 F_{\text{LF}}
\cdot
 \,2^{-\text{C}}
 \cdot  
 (\kappa_p + \kappa_i\frac{z^{-1}}{1-z^{-1}})
  \cdot 
 (\frac{z^{-1}}{1-z^{-1}})
 \cdot  
 z^{-\text{D}} \cdot 
\label{TF_G_complete}
\end{equation}
The system transfer function $H(z)$ and the error function $E(z) = 1 - H(z)$, which describes the untracked parts of the input signal and therefore the tracking error, are derived from the open-loop transfer function.
\begin{equation}
		H(z) = \frac{G(z)}{1+G(z)}= \frac{\varepsilon_o}{\varepsilon_i},
	\label{TF_H}
\end{equation}
\begin{equation}
		E(z) = \frac{1}{1+G(z)}= \frac{\varepsilon_e}{\varepsilon_i}.
	\label{TF_E}
\end{equation}
We performed a loop gain measurement of one of our VHDL implementations by adding digital noise into the loop. The results in figure \ref{figure3} show a very good agreement between the implementation and the simulation, with the exception of the peak corresponding to the second harmonic frequency (20MHz).
\bfseries{One should keep in mind though that this model and all corresponding analysis is only valid in closed loop operations and only if the gain of the loop is sufficient to maintain the error point close to zero.}\mdseries

\subsection{Additive and phase noise}

We now extend the input signal (equation \ref{incoming1}) by including additive noise $\widetilde{A}$ (including shot noise, electronic noise and relative intensity noise) and phase noise $\widetilde{\varepsilon}$. Since the PLL can not distinguish between phase noise and phase signal $\varepsilon_{s}$ both terms can again be described by a single term $\varepsilon_{i}$.
\begin{equation}
		i[n] = \widetilde{A}[n] + A \cdot \sin{\left( \omega_0 n  + \widetilde{\varepsilon_{i}} + \varepsilon_{s}[n] \right)} = \widetilde{A}[n] + A \cdot \sin{\left( \omega_0 n  + \varepsilon_{i}[n] \right)}. 
	\label{eq:incoming_noise}
\end{equation}
For phase noise the standard deviation of the residuals $\sigma_{\text{phase}}$ can be
computed by integrating the product of the noise with the loop
error function $E(z)$: 
\begin{equation}
\sigma^2_{\text{phase}} = \int_0^{\infty}{\widetilde{\varepsilon}_{i}^2(z) \times E(z)^2 df }
 	\label{eq:freqvar}
\end{equation}
This is a measure of untracked residual phase error. 

The standard deviation of the error generated by input additive noise $\sigma_{\text{add}}$  is computed by
integrating the product of the effective phase noise with the system transfer
function $H(z)$, since this transfer function describes how 
noise added to the error signal $\varepsilon_{e}(z)$ is attenuated in
a closed loop. Due to the mixing process the amplitude noise induced phase noise $\widetilde{\varepsilon}_{\textrm{amp}}$ is also increased by $\sqrt{2}$ \cite{Sheard2011}
\begin{equation}
\sigma^2_{\text{add}} = \int_0^{\infty}{  \left( \frac{\sqrt{2} \widetilde{A}}{A} \right)   ^2(z) \times H^2(z) df } =
 \int_0^{\infty}{  \left( \widetilde{\varepsilon}_{\textrm{add}} \right)   ^2(z) \times H^2(z) df }.
 	\label{eq:shotvar}
\end{equation}
The different treatment of phase noise and additive noise, both
of which are present in the input signal, can be understood if one considers phase as the
quantity that propagates around the loop. Input phase noise directly represents that and an increased
bandwidth allows to track this phase more precisely. Additive noise as such does
not represent a phase error. It only gets converted into phase
noise by the action of the mixer (phase detector) which is {\em in
the loop}, hence the different transfer function.

\subsection{Amplitude detection}

The detection of the amplitude A of the incoming signal is performed 
by multiplying $i[n]$ with an in-phase output of the NCO $I[n]$. 
\begin{equation}
\begin{split}
		I[n] &= \frac{1}{2} \cdot \sin{\left( \omega_0 n  + \varepsilon_{o}[n] \right)}, \\
		\left[I[n] \right] &= 1. 
	\end{split}
	\label{eq:ampdec}
\end{equation}
The multiplication of $i[n]$ and $I[n]$ gives

\begin{equation}
	u_I [n] = 	 \frac{A}{4} \cdot \big[  \cos{\left(  \varepsilon_{i}[n] - \varepsilon_{o}[n] \right)}  - 
		 \cos{\left(  2 \omega_0 n  + \varepsilon_{i}[n] + \varepsilon_{o}[n] \right)}     \big].
	\label{eq:ampl}
\end{equation}
Assuming a locked PLL $(\varepsilon_{i} - \varepsilon_{o} = \varepsilon_{e} << 1)$ 
and a sufficient filtering of the second harmonic, this can be reduced to

\begin{equation}
	u_I [n] = 	 \frac{A}{4} \cos \left(  \varepsilon_{e}[n] \right) \approx \frac{A}{4}.
	\label{eq:redampl}
\end{equation}
Therefore the readout of $u_I$ (I) yields directly the amplitude $\frac{A}{4}$ of the tracked tone,
while DC offsets and signals at sufficiently different frequencies average to zero. 
Knowledge of the signal amplitude is required to understand the loop bandwidth, to track changes in the interferometry,
like in contrast and optical power, and to perform calculations involving the vector properties of the input signal, like for example
stray light corrections \cite{Dehne2012}.

\section{Readout}
\subsection{Frequency readout}
The phase $\varepsilon_{i}$ can be reconstructed by 
reading the frequency value $u_f$ (PIR) or the phase value $u_p$ (PA) 
of the PLL, which represent the 
frequency/phase of the incoming signal, respectively. 
For the PA:

\begin{equation}
\begin{split}
	u_p[n] & \approx (\omega_0n+ \varepsilon_i[n])/ 2\pi ~~~ (\text{for}~\varepsilon_{e}[n] \ll 1   ).\\
		 \left[u_p[n] \right] &= \text{cycle}; ~ \left(-\pi~\text{rad} \leq (2 \pi \times u_p[n]) < \pi~\text{rad} \right).
	\end{split}
	\label{eq:readout:pa}
\end{equation}
Since the absolute system phase is a ramp, with the slope given by the current heterodyne frequency, 
a direct readout of $u_p$ is not practical since this value will overflow very quickly. A decimation
of such a sawtooth function is difficult, and the dynamic range for a non overflowing value of $u_p$
is very large.
The preferred possibility for the phase readout of a single loop is the frequency value $u_f$. 
\begin{equation}
\begin{split}
	u_f[n] &\approx ( \omega_0 + \frac{\delta \varepsilon_i}{\delta \tau_s} )/ 2\pi ~~~ (\text{for}~\varepsilon_{e}[n] \ll 1   ).\\
		 \left[u_f[n] \right] &= \frac{\text{cycle}}{\text{s}}; ~ \left(0~\text{Hz} \leq (f_s \times u_f[n]) < f_s \right).
	\end{split}
	\label{eq:readout:pir}
\end{equation}
This value
is not overflowing and allows for standard decimation and filtering algorithms to be implemented, though one has to keep in mind that this
signal has a large dynamic range. Any requirements on decimation filters and bit length have to take into account that the signal of interest (phase) is not directly processed, but its derivative, which changes its spectral properties. The phase fluctuations can easily be reconstructed afterwards by integration.

\subsection{PA readout}
If several channels track the same frequency and they only vary slightly in phase, the differences of the PLL phases $(\Delta u_p)$ can be read out directly by subtracting the PA values. The rapid ramp present in the individual loops is thereby completely subtracted and only the small signal of interest remains.
This is ideal for implementing techniques like differential wave front sensing (DWS) \cite{Schuldt2009}. 

Even tough the small differences in phase can also be reconstructed from the PIR readout, the PA readout is preferred. This is because the PIR values need to be tracked continuously to reconstruct the correct absolute phase values. This means that any glitches or cycle slips will break the reconstruction. Even though the reconstruction can be restarted, a new initialization of the PLLs would be required. In contrast to that the PA readout does not break by such an event, (assuming in both cases that the PLLs stay in lock) but it would automatically return to the correct value after the event passed, making it more reliable. In addition the PA difference signals have a much smaller dynamic range, allowing to reduce the required bit lengths and computational efforts in further processing of, for example, DWS data.



\subsection{Additional IQ readout}
If the residual phase error in the loop
$\varepsilon_{e}$ exceeds the acceptable noise level,
because the loop reacts too slowly to track precisely the 
phase fluctuations of the incoming signal ($RMS[\varepsilon_e(>1~\textrm{Hz})]>1~\mu \textrm{cycle} $ for LISA),
an additional corrective readout can be performed \cite{Shaddock2006}. This might be necessary if the required PLL bandwidth needs to be
rather low to achieve stable operations (see section \ref{sec5}).

Since the untracked signal in a PLL is a vector and not a scalar,
the readout of both quadrature components I and Q is required for 
the additional phase reconstruction. For a loop locked near but not exactly on
zero phase difference, they can be written as

\begin{equation}
\begin{split}
	Q = u_e [n] =& 	\, \frac{A}{4} \sin{\left(  \varepsilon_{e}[n]  \right)} \\
	I = u_I [n] =& 	\, \frac{A}{4} \cos{\left(  \varepsilon_{e}[n]  \right)}. 
	\end{split}
	\label{eq:quadcomp}
\end{equation}
The residual phase $ \varepsilon_{e} $ can be reconstructed by computing

\begin{equation}
	\varepsilon_{e} = \arctan {\left(  \frac{u_e [n]}{u_I [n] }\right)} = \arctan {\left(  \frac{Q}{I }\right)}.
	\label{eq:arctan}
\end{equation}
Equally the amplitude A 
of the vector in this situation can be computed as

\begin{equation}
	\frac{A}{4} = \sqrt{ {u_e [n]}^2+ {u_I [n]} ^2}      = \sqrt{Q^2+I^2}.
	\label{eq:squareamp}
\end{equation}
Which readout is required can be evaluated by comparing the
PLL bandwidth with the dynamic range of the incoming signal. For the design shown here we used a controller that has sufficient signal suppression at low frequencies to reach the required performance without additional IQ readout. Nevertheless we still implemented it for diagnostic purposes.

\subsection{Decimation}

The signals of interest are decimated to a desired sampling rate (typically of the order of a few Hz) for storage and further computation. 
The decimation can be implemented in one or several steps and can make use of different computation methods, based on the hardware used. Here we only describe the decimation taking place inside the FPGAs, which is normally restricted to use integer parallel processing.

We found CIC filters \cite{Hogenhauer1981} to be a good choice for the decimation inside FPGAs. Their implementation is simple (they only require accumulators and differentiators), they are easily modelled and they provide notches of suppression exactly at the most critical frequencies, the ones that would be aliased to very low frequencies. Which order of filter is required can be computed for each signal by comparing the sum of all frequencies filtered and aliased into the signal band to the requirements. Since the suppression of CIC filters is increasing with frequency this calculation is, in the case of LISA like signals, completely dominated by the first notch. 

The use of CIC filters also allowed us to implement an additional noise shaping technique \cite{Lehtinen2002}. This technique allows to reduce the readout bit length of some signals, since it reduces low-frequency truncation noise due to the CIC transfer functions. 

\section{Quantization Noise Model}
\subsection{Truncation Noise}
The digital integer numbers used in the FPGA implementation can only 
represent a finite number of
distinct values with a constant, non-zero separation between them.
An effective implementation of an ADPLL will require the use of signal truncations inside
the loop, to save resources and to fit into flight compatible FPGA devices.
The modelling and implementation of such truncations is described in the following.

It is well known that the truncation of a continuous signal to a digital number 
with $N$ bits at a sampling
rate $f_{\rm samp}$ can be modelled as an addition of uniformly distributed
white noise with a linear power spectral density of 
\begin{equation}
\begin{split}
		\widetilde{x}_{\rm trunc} 
		&= \frac{q}{ \sqrt{6\cdot f_{\rm samp}}}
		= \frac{2^{-N}}{ \sqrt{6\cdot f_{\rm samp}}}. \\
		\left[\widetilde{x}_{\rm trunc}\right] &= \frac{1}{\whz}.
	\end{split}
	\label{DigNoise}
\end{equation}
The same formula is also applied here for the truncation of digital signals,
though the assumption of additive white noise is only valid for signals that 
move through a significant range
of digital values without any coherent relationship to the sampling frequency,
like e.g.\ two or more sine waves at non-harmonic
frequencies \cite{Widrow2008}. Otherwise the quantized signal will show
artefacts and peaks from the coherent interaction with the truncation process. 

\subsection{Dither}
To avoid such artefacts, an intentional noise floor is added to the signal before truncation,
so called {\bf dither}, with triangular dither being the preferred implementation \cite{Widrow2008}.

Such a triangular dither generator was implemented by subtracting the outputs of two independent linear feedback shift registers with a repetition length longer than 10000\,s, to ensure that no artefacts will be visible in the LISA signal spectrum ($0.1\,\textrm{mHz}$ to $1\,\textrm{Hz}$).

Based on simulation the effective white noise introduced by a dithered truncation was found to be slightly higher with a value of
\begin{equation}
\begin{split}
		\widetilde{x}_{\rm trunc+dith} &= \frac{2^{-N}\sqrt{3}}{\sqrt{6\cdot f_{\rm samp}}}= \frac{q\sqrt{3}}{\sqrt{6\cdot f_{\rm samp}}}. \\
		\left[\widetilde{x}_{\rm trunc}\right] &= \frac{1}{\whz}.
	\end{split}
	\label{DigNoise2}
\end{equation}
The increase by a factor of $\sqrt{3}$ can be tolerated, since the introduction of spurious signals is now suppressed and anyway this noise can be arbitrarily reduced by using more bits. 

\subsection{Rounding}
Truncation can also introduce 
small signal offsets due to rounding errors. This is prevented by offset-free rounding algorithms based on simple integer arithmetic. We designed specific VHDL rounding blocks for our implementation. These blocks truncate symmetrically around zero, and they are linear, keeping the amplitudes of signals constant. To generate the correct offset for some truncation cases a dithered bit is used to determine the rounding direction.

\subsection{Noise shaping}
The linearised ADPLL model can now be used to understand the effect of in-loop truncation noise on the phase tracking performance, by applying standard control theory. 
As an example we evaluate the influence of a truncation of the frequency value $u_f(z)$,
it shows a strong noise shaping
and it also allows a strong reduction of readout bit rates. 

A naive PLL implementation would use a high number of bits at $u_f(z)$. 
This is because truncations at this point in an open-loop system are especially critical, 
since they introduce a white frequency noise, which leads to a 
$1/f$ phase noise inside the PLL. The linear model shows, however, that this noise is suppressed directly by the loop
error function $E(z)$. Since the PLL includes a $f^2$ suppression at low frequencies, the effective phase noise is easily
reduced below $1\mu\textrm{rad} / \sqrt{ \textrm{Hz}}$ in the LISA signal bandwidth.
\begin{equation}
		\widetilde{\varepsilon}_{u_{f}} (z) = \sqrt{3}~ \frac{f_s}{f} \frac{2^{-T}}{\sqrt{6\cdot f_{s}}} * E(z) ~ \text{rad}.
	\label{freqtruncnoise}
\end{equation}
We implemented such a truncation and were able to perform null measurements with a performance of $1\,\mu\text{cycle}$ between $0.1\,\textrm{mHz}$ and $1\,\textrm{Hz}$ in a $80 \textrm{MHz}$ system with a 
12 bit frequency value,
which corresponds to an LSB frequency resolution
of only $\approx 20~\textrm{kHz}$. All of the following simulations include this truncation.

The bit reduction of this specific signal is especially useful, since the readout of the PLL frequency
requires the highest dynamic range of all PLL signals. The noise shaping allows to reduce the initial bits to be downsampled and potentially also reduces all bit lengths in further processing.

\section{Non-linearity and cycle slips}
\label{sec5}
The presented linear analysis of the phasemeter is valid for many applications. Phasemeters based on this
analysis were already successfully tested and used in various laboratory experiments \cite{Mitryk2010,Spero2011}.


A phasemeter in a true inter-satellite interferometer
will have to operate under rather extreme conditions, a low SNR of the input signal (due to additive noise) and a large dynamic range of the signal phase, due to frequency noise and signal dynamics. Therefore the PLL could reach a state where it becomes non-linear and
cycle slips occur in the PLL tracking \cite{Ascheid1982,Viterbi1963}.

The resulting phase noise from $R$ slips during a measurement period is given as \cite{Dick2011}
\begin{equation}
\phi_{slip} (z) = \frac{\sqrt{2} \sqrt{R}}{f} ~ \text{rad}.
 	\label{slipnoise}
\end{equation}
For LISA this means that any slip spoils the system performance completely and
is therefore comparable to a loss of lock or another measurement disturbance. 
Earlier experimental investigations by Dick et. al. \cite{Dick2011} and detailed modelling \cite{Ascheid1982,Viterbi1963} have shown that the relation between the bandwidth and the signal noise floor 
is the critical factor for the probability of cycle slips. 

Since the LISA Phasemeter needs to operate far outside any cycle slip region and therefore
in the linear regime, we compiled a model to determine a suitable loop bandwidth for a given set of signal parameters, that should allow to minimize the cycle-slip probability and non-linear effects of the phase tracking.

The two most important reasons for non-linear behaviour 
are the sinusoidal response of the phase detector and the 
existence of second harmonics and other additional tones, like the side-band beatnotes
for inter spacecraft clock transfer or an ADC pilot tone.

\subsection{Phasedetector}

The non-linear output of the phase detector, omitting the second harmonic, is 
\begin{equation}
	u_e [n] = 	 \frac{A}{4} \cdot \sin{\left(  \varepsilon_{i}[n] - \varepsilon_{o}[n] \right)}= 	 \frac{A}{4} \cdot \sin{\left(  \varepsilon_{e}[n] \right)}.
	\label{mixer4}
\end{equation}
The non linear response is also shown in the inset of figure \ref{figure4} in
direct comparison to the linear behaviour assumed before. 

For an error signal $\varepsilon_{e}[n] \ll
\pi/2$ we can assume a quasi-linear behaviour of the phase detector. If the error signal exceeds $\pi/2$ the
loop gain starts to reduce until it crosses zero and changes sign at $\varepsilon_{e}[n] =
\pi$. At any of these points the loop is potentially unstable and the error signal
eventually jumps by $2\pi$ or more, which results in a phase tracking
error of the same amount. 

\subsection{Optimal bandwidth}
Since the absolute error signal is directly related to the linearity of the PLL we can use the calculated standard deviations to evaluate the size of the error signal. We also include the second harmonic and digitisation noise from inside the PLL to get a complete picture. By evaluating this for different loop bandwidths we can then optimize the PLL to work in a regime closest to linear behaviour, minimizing non-linear effects and cycle slips.

The standard deviations for additive and phase noise have already been calculated in section 1.
The internal quantization noise influences can be added
quadratically (assuming uncorrelated noise sources). The
comparison of the resulting standard deviations can be compared
for different bandwidth and phase margin (damping) configurations
to find an optimal design of the PLL. The standard deviations can be added quadratically to compute the
resulting overall standard deviation $\sigma_{sum}$. 

\begin{figure}
   \centering
   	\includegraphics[width=0.80\textwidth]{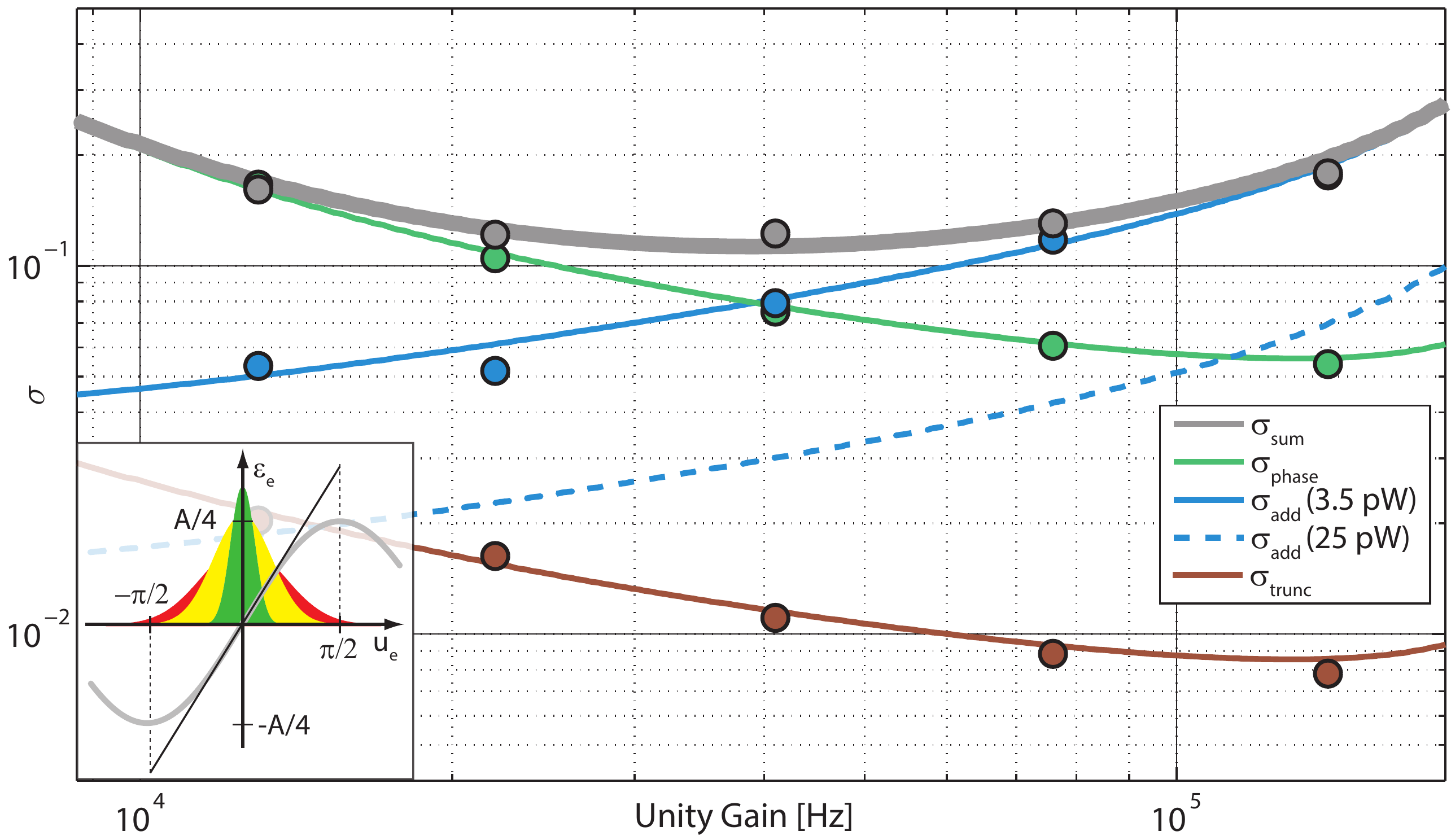}
   		\caption[figure4]{\label{figure4}
	Shown is the $1~\sigma$ standard deviation of the error point from additive noise (blue), phase noise (green), truncation noise (brown) and of their quadratic sum (grey). The additive noise is an example based on the laser shot noise expected in LISA with an effective received power of $3.5\,\textrm{pW}$ (also shown for comparison is the additive noise for $25\,\textrm{pW}$ effective power). The phase noise used here is the laser frequency noise expected in LISA (see $\widetilde{\epsilon_i}$ in figure \ref{figure5}).  
The dots show the measured values of $\sigma$.
	
The inset is a schematic of the linearised and real response of the phase detector in comparison to different distributions of the error point signal. The green distributions illustrates a linear case, the yellow and the red curves show how the non-linearity becomes more prominent as $\sigma$ increases. }
\end{figure}

Figure \ref{figure4}
shows the modelled standard deviations for example parameters (discussed in section \ref{sec6}) and their dependency on the loop gain. To verify our model we have measured the standard deviation of the PLL error signals for various noise influences and a range of stable bandwidth. We computed the standard deviations by fitting the phase error of the PLL, which we read out at full sampling speed by subtracting the PA value of the PLL and an NCO used in our simulations. The measured standard deviations are shown as dots in figure \ref{figure4} together with their respective modelled values. Our model shows excellent agreement between the predictions and the measured values, which verifies that the linear model is appropriate for this range of operation. 

The optimal bandwidth for the here assumed noise sources is found at $\approx 40\,\textrm{kHz}$. Operating the PLL at this point should allow to minimize any non-linear phase artefacts and the probability of cycle slips.
Though we can not deduce the exact probability, we can now test the system for stability and performance for given signal parameters.

\subsection{Second harmonic}
The second non-linear behaviour of the phase detector is the generation of a
second harmonic of the input signal, as shown in equation \ref{eq:mixer1}.

We can split the effects of the second harmonic into two parts. 
The first effect creates parasitic phase noise in the signal band, which we describe in detail in the following. 
For convenience we therefore rewrite the second harmonic part of equation \ref{eq:mixer1} in the continuous time domain
\begin{equation}
	u_{e,2f} (t) = 	 \frac{A}{4} \cdot \sin{\left(  2 \omega_0 t  + \varepsilon_{i}(t) + \varepsilon_{o}(t) \right)}.
	\label{secondharmonic}
\end{equation}
We simplify this equation by assuming the PLL to be tightly locked $(\varepsilon_o \approx \varepsilon_i)$ and by defining an effective phase value $\varepsilon_{\textrm{eff}}(t) = \omega_0 t + \varepsilon_{o}(t)$, with an effective frequency  $\omega_{eff} = \frac{\delta \varepsilon_{\textrm{eff}}}{ \delta t}$.
The second harmonic propagates through the PLL in a time $\tau_g$ and creates an effective phase modulation.
The output of the NCO at the time $t$ can therefore be written as
\begin{equation}
	\textrm{NCO}_{\textrm{out}} (t) = 	 \frac{1}{2} \cos \Big( \varepsilon_{\textrm{eff}}(t) +
	m \cdot \sin{(  2 \varepsilon_{\textrm{eff}}(t-\tau_g) )} \Big).
	\label{secondharmonic3}
\end{equation}
Here $m$ is a modulation index given by the attenuation of the second harmonic by the open loop transfer function ($m = |G(2 \omega_{\textrm{eff}})|$), referred to the gain for the nominal low-frequency error signal, which in the signal range is $\approx 1$.
Using Bessel functions of the first kind we can expand this to 
\begin{equation}
\begin{split}
	\textrm{NCO}_{\textrm{out}} (t) = 	 &\frac{1}{2} J_0(m) \cos ( \varepsilon_{\textrm{eff}}(t) ) + 
	\frac{1}{2} J_1(m) \sin ( \varepsilon_{\textrm{eff}}(t) )  \sin ( 2\varepsilon_{\textrm{eff}}(t-\tau_g) )  + \mathcal{O}(m^2) \\
	= &o(t) + o_{2f}(t) + \mathcal{O}(m^2) .
	\end{split}
	\label{secondharmonic4}
\end{equation}
Assuming $m \ll 1$ one can approximate the first two Bessel function by $J_0(m) \approx 1$ and $J_1(m) \approx m/2$. This yields the original NCO output $o(t)$, the term from the second harmonic $o_{2f}(t)$ and higher terms $ \mathcal{O}(m^2)$, which we discard in the following. We now rewrite $o_{2f}$ and immediately discard the third harmonic term. 
\begin{equation}
\begin{split}
	o_{2f}(t) = &	 \frac{1}{2} \frac{m}{4} \big( \cos ( \varepsilon_{\textrm{eff}}(t-\tau_g)) -
	  \cos ( 3\varepsilon_{\textrm{eff}}(t-\tau_g)  )   \big)  \\
	  	o_{2f}(t) \approx &	 \frac{1}{2} \frac{m}{4}  \cos ( \varepsilon_{\textrm{eff}}(t-\tau_g) ) 
	  \end{split}
	\label{secondharmonic5}
\end{equation}
The phase modulation side-band at $\omega - 2 \omega = -\omega$ thus ends up at the same frequency as the nominal NCO output and results in a parasitic phase signal by the action of the mixer.
The additional mixer output is
\begin{equation}
	i(t) \times o_{2f}(t) \approx  \frac{A}{4} \frac{m}{4}  \sin \big( \varepsilon_{\textrm{eff}}(t) - \varepsilon_{\textrm{eff}}(t -\tau_g) \big).
	\label{secondharmonic6}
\end{equation}
The effective parasitic phase error $\varepsilon_{p,2f}$ is therefore (see equation \ref{mixer2})
\begin{equation}
	\varepsilon_{p,2f} (t) = \frac{m}{4}  \sin \big( \varepsilon_{\textrm{eff}}(t) - \varepsilon_{\textrm{eff}}(t -\tau_g) \big).
	\label{secondharmonic7}
\end{equation}
Assuming a constant $\tau_g$ and an effective frequency $\omega_{\textrm{eff}}$ that 
varies on time-scales smaller than $\tau_g$ one can approximate this to
\begin{equation}
	\varepsilon_{p,2f} (t) \approx \frac{|G(2 \omega_{\textrm{eff}})|}{4}  \sin \big( \omega_{\textrm{eff}}(t) \tau_g \big) .
	\label{secondharmonic8}
\end{equation}
This parasitic noise couples very non-linear and depends highly on the suppression of the second harmonic $m$ , the group delay in the PLL $\tau_g$ and the dynamics of the input signal $\omega_{\textrm{eff}}(t)$. The coupling is at its maximum in the linear range of the sine. Assuming this operating condition we can calculate the maximum parasitic phase error in dependency of the signal frequency noise spectrum.
\begin{equation}
	\widetilde{\varepsilon}_{p,2f} (f) \approx \frac{|G(2 \omega_{\textrm{eff}})| }{4}  \, \widetilde{\omega}_{\textrm{eff}}(f) \tau_g .
	\label{secondharmonic9}
\end{equation}
The second effect caused by the second harmonic is an additional instantaneous root-mean square value of the error signal. Even though this does not cause a phase error at low frequencies, it does increase the probability to leave the linear range of the phase detector. The maximum additional error is $\varepsilon_{e,2f} (max) = |G(2 \omega_{\textrm{eff}})|$. 

The above equations allow to determine the necessary suppression by low pass filters for a given system, by calculating the error signal residuals and by comparing the signal dynamics with the required phase performance.
Since the choice of low pass filter is also limited by logic resources, a trade-off is necessary. We have found IIR filters to be a good compromise between suppression and logic resources required. A second order IIR filter with a corner frequency of $300\,$kHz is used in the following simulations, a small residual parasitic phase is visible as the roll-up in the blue curve in figure \ref{figure5}. In critical cases, e.g. when the signal frequency can span a wide range, a more complex $2f$-filter could be used, for example one that adapts its corner frequency to the signal frequency. One should also consider that this analysis is only valid if the second harmonic is below the Nyquist frequency ($f_s/2$). If this is not the case the second harmonic will be aliased to another frequency and potentially not cause a parasitic phase error.

\section{Digital measurements}
\label{sec6}

\begin{figure}
   \centering
   	\includegraphics[width=0.85\textwidth]{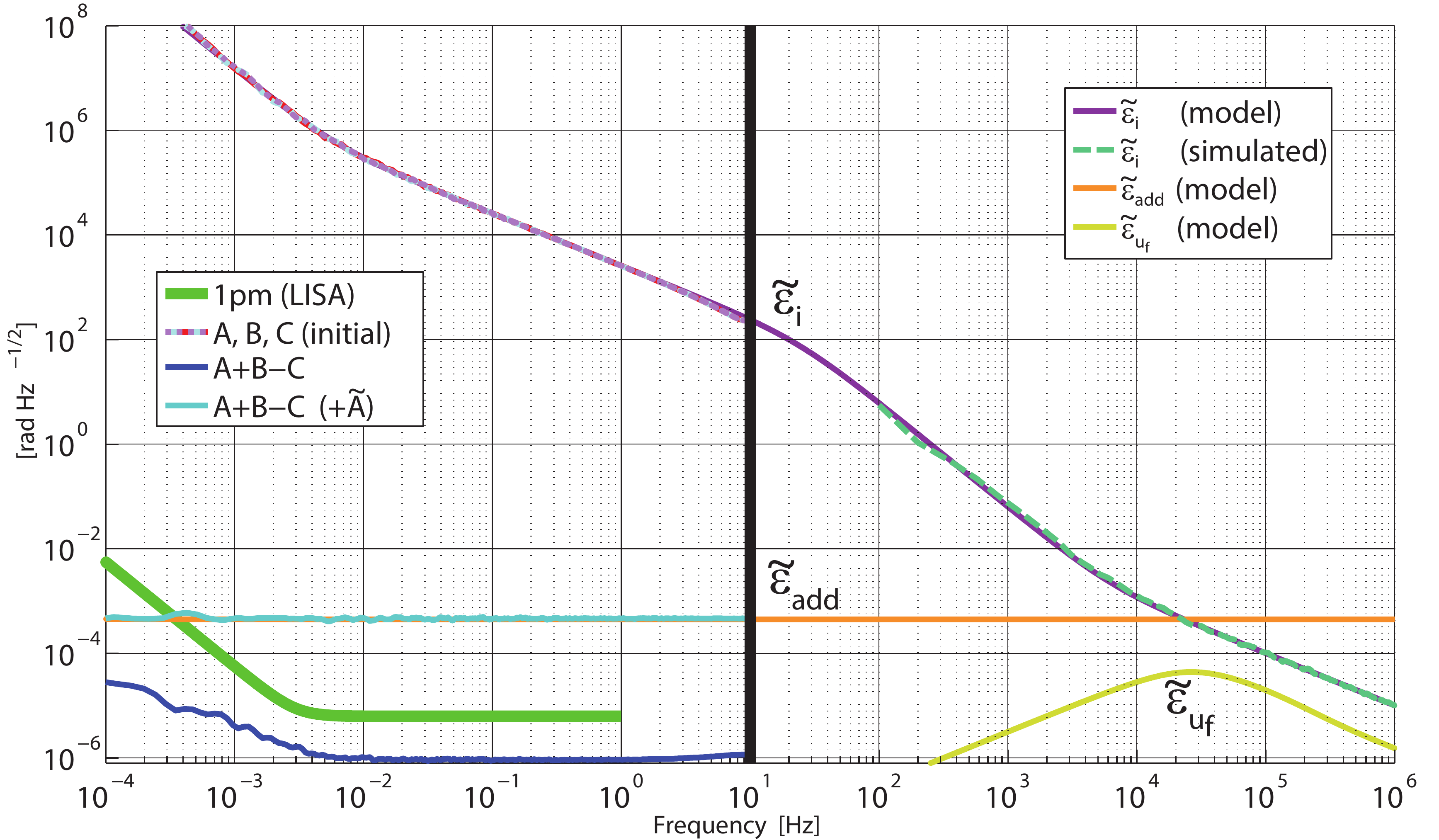}
   		\caption[figure5]{\label{figure5}
   		(left) Shown are the results of two digital non-linearity measurements. The initially measured signals A, B and C are the same for both measurements and are therefore only plotted once. The first measurement (dark blue) was performed without any additive noise, the correct combination of the input signals reveals the noise floor and linearity limits of the PLLs under test. It demonstrates the full performance of the PLL, only limited by numerical limits.  The measurement also shows a dynamic range performance of the phasemeter of 10 orders of magnitude at $0.01\,\textrm{Hz}$, necessary for the implementation of TDI. We observe a small roll-up at low frequencies, which we attribute to truncations in data post-processing and non-linearities in the PLL due to the second harmonic.
The second measurement (light blue) used additional additive noise in all three signals with a SNR equivalent to $3.5\,$pW effective power in a LISA-like set-up. No cycle slips were observed under these extreme conditions and correct signal combination reveals the performance to be limited by a white noise floor. \\
   		(right) Shown here is the high-frequency part of the phase noise used for the signal (a model in violet and a simulation in dashed blue), as well as the predicted noise floor for the additive noise (orange) and the expected phase noise due to the truncation to 12 bit at the PIR (yellow).
		}
\end{figure}

To evaluate the modelled performance and noise influence, we performed FPGA based
measurements of the ADPLL performance. Similar to Shaddock et al. \cite{Shaddock2006} we implemented a scheme based on a digital non-linearity
test, where three independent noise sources are generated, combined
and then fed into three numerically controlled oscillators. Tracking all three signals
and combining their respective phase measurements allows to determine the phase noise performance
for large signals, and under realistic conditions. 

A white gaussian noise, generated like the truncation dither, is shaped by a specially designed IIR filter, to simulate the laser frequency noise expected at the beatnote of the master satellite in the LISA configuration. For our signals we choose the highest pre stabilised laser frequency noise spectrum proposed for LISA (800$\,\textrm{Hz}/\sqrt{\textrm{Hz}}$ in band) \cite{Sheard2011}.

An additional gaussian noise is used to introduce additive noise and to simulate a weak light environment, here with an effective power of $\approx 3.5\text{pW}$, corresponding to an SNR of $30 \, \textrm{dBHz}$.

The PLLs used to track these three beatnotes are optimised based on the described
models and techniques.
 This includes the frequency truncation to 12 bits, the readout truncation, loop gain optimisation and
 sufficient second harmonic filtering.
 
We performed two of these measurements, one with weak light condition and one without, to test the PLL stability and the performance. The result of both are shown in figure \ref{figure5}. 
 
For the weak light case the measurement shows a continuous tracking of all signals, without the occurrence of cycle slips. The achieved performance after signal recombination was limited by the additive noise, as expected.

Without additional noise the measurement achieved a performance better then $1\,\mu\text{cycle}/\sqrt{\text{Hz}}*NSF$ in the whole signal range. This demonstrates that the underlying noise floor of the system is sufficient for LISA-like missions. We could thereby demonstrate a dynamic range of up to $10^{12}$ at $1 \, \textrm{mHz}$.

\section{Analog measurements}

\begin{figure}
   \centering
   	\includegraphics[width=0.95\textwidth]{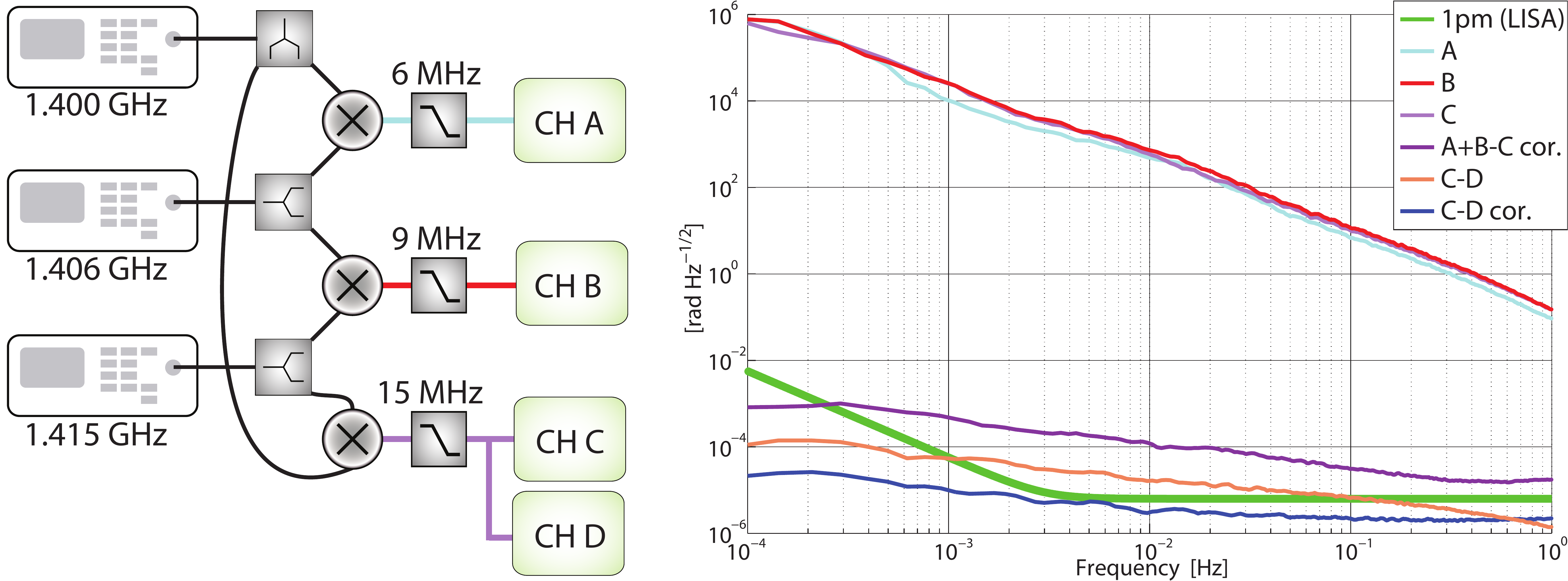}
   		\caption[figure6]{\label{figure6}
(left) Shown is the set-up used for the analog three signal test, generating three signals with phases that can be combined to zero.\\
(right) Shown are the measurement results achieved with this set-up. One of the signals is split after the mixing and fed into two channels to investigate the noise floor of the phasemeter prototype used. The initial noise floor of this reference measurement (orange) lies above the requirement for a wide range of the spectrum. The use of a pilot-tone correction allowed to reduce this noise below the requirements for the full range (dark blue). The readout of this null measurement was performed using the readout of the PIR and the PA (not shown), both results are indistinguishable in the required frequency range and show only slight variations at high frequencies, due to different transfer functions and aliasing.
The three signal combination (violet) reveals a noise floor above the requirement for almost all frequencies, not allowing us to fully test the linearity of our phasemeter channels. The cause of this excess noise was identified to be the mixers, that generate a low frequency phase noise that spoils the performance.}
\end{figure}

The three signal test was also performed using analog signals. By mixing three GHz tones, we generated three MHz signals with similar properties as in the simulations. Those three signals were injected into a phasemeter prototype \cite{Shaddock2006} and the measurement signals and combinations are shown on the right side of figure \ref{figure6}. The analog mixing is limited by low frequency phase noise generated in the mixers and can therefore not show the full system performance. The phasemeter noise performance, including digitisation noise and analog front-end noise, was demonstrated in parallel by a null measurement. The use of a pilot tone allowed to correct this measurement below the LISA requirement, showing that the front-end in the experiment performed as required.

Though the full performance was not yet shown with analog signals, we already reached a dynamic range of up to $10^7$ at $1\,\textrm{mHz}$. 

\section{Conclusion}

We have demonstrated a full model of the phase readout system for future LISA like space borne gravity missions. 
We have used this model to design and optimize the system parameters and to predict the influence of truncations.
Non-linearities were treated in three steps, first by applying the linear model to find the optimal bandwidth, second by testing the designed PLL in a realistic VHDL based measurement and third by using real analog signals with similar properties.

Future plans include the testing of the phasemeter performance with analog and optical signals, to perform tests under more realistic conditions and to include further noise influences. An interesting idea for future work might be to further investigate the ratio between the standard deviation and the cycle slip probability. An automatic loop gain control will also potentially be necessary to stay in the linear system range. The phasemeter core will also be adapted for the use in an Breadboard Model of the LISA phasemeter currently built and tested in an ESA technology development activity \cite{Gerberding2012}.

\section{Acknowledgements}
We acknowledge funding by the European Space Agency (ESA) within the technology development project ``LISA Metrology System"´. The authors also gratefully acknowledge support by the International Max-Planck Research School for Gravitational Wave Astronomy (IMPRS-GW), by the Centre for Quantum Engineering and Space-Time Research (QUEST) and by Deutsches Zentrum f\"ur Luft- und Raumfahrt (DLR) with funding from the Bundesministerium für Wirtschaft und Technologie (project reference 50 OQ 0601).

\end{document}